\documentclass[apj]{emulateapj}

\newcommand{\kms}{km~s$^{-1}$}

\newcommand{\mg}{\textrm{Mg}}
\newcommand{\fe}{\textrm{Fe}}
\newcommand{\mgfe}{[\textrm{Mg}/\textrm{Fe}]}
\newcommand{\mgii}{\textrm{Mg}~\textsc{ii}}

\newcommand{\hi}{\textrm{H}~\textsc{i}}

\newcommand{\feii}{\textrm{Fe}~\textsc{ii}}

\newcommand{\oii}{[\textrm{O}~\textsc{ii}]}
\newcommand{\oiii}{[\textrm{O}~\textsc{iii}]}
\newcommand{\lobs}{\lambda_{\textnormal{\scriptsize{obs}}}}
\newcommand{\lrest}{\lambda_{\textnormal{\scriptsize{rest}}}}

\newcommand{\mstar}{\ensuremath{\mathcal{M}_{*}}}
\newcommand{\msun}{\ensuremath{\mathcal{M}_{\odot}}}
\newcommand{\sigmasfr}{\Sigma_{\textnormal{\scriptsize{SFR}}}}

\shorttitle{Co-Rotating Halo Gas}
\shortauthors{Diamond-Stanic et al.}

\begin{document}
\slugcomment{Submitted to ApJ}

\title{ Galaxies Probing Galaxies at High Resolution: \\ Co-Rotating
  Gas Associated with a Milky Way Analog at z=0.4}

\author{Aleksandar M. Diamond-Stanic\altaffilmark{1,2}, Alison
  L. Coil\altaffilmark{3}, John Moustakas\altaffilmark{4}, Christy
  A. Tremonti\altaffilmark{1}, Paul H. Sell\altaffilmark{5}, Alexander
  J. Mendez\altaffilmark{6}, Ryan C. Hickox\altaffilmark{7}, \& Greg
  H. Rudnick\altaffilmark{8}}

\altaffiltext{1}{Department of Astronomy, University of
  Wisconsin-Madison, Madison, WI 53706, USA}
\altaffiltext{2}{Grainger Fellow; aleks@astro.wisc.edu}
\altaffiltext{3}{Center for Astrophysics and Space Sciences,
  University of California, San Diego, La Jolla, CA 92093, USA}
\altaffiltext{4}{Department of Physics and Astronomy, Siena College,
  Loudonville, NY 12211, USA}
\altaffiltext{5}{Department of Physics, Texas Tech University,
  Lubbock, TX 79409, USA}
\altaffiltext{6}{Department of Physics and Astronomy, Johns Hopkins
  University, Baltimore, MD 21218, USA}
\altaffiltext{7}{Department of Physics and Astronomy, Dartmouth
  College, Hanover, NH 03755, USA}
\altaffiltext{8}{Department of Physics and Astronomy, University of
  Kansas, Lawrence, KS 66045, USA}

\begin{abstract}

  We present results on gas flows in the halo of a Milky Way--like
  galaxy at $z=0.413$ based on high-resolution spectroscopy of a
  background galaxy.  This is the first study of circumgalactic gas at
  high spectral resolution towards an extended background source
  (i.e., a galaxy rather than a quasar).  Using longslit spectroscopy
  of the foreground galaxy, we observe spatially extended H$\alpha$
  emission with circular rotation velocity $v_{circ}\approx270$~\kms.
  Using echelle spectroscopy of the background galaxy, we detect
  \mgii\ and \feii\ absorption lines at impact parameter $\rho=27$~kpc
  that are blueshifted from systemic in the sense of the foreground
  galaxy's rotation.  The strongest absorber
  ($\textnormal{EW}_{2796}=0.90$~\AA) has an estimated column density
  ($N_H\geq10^{19}$~cm$^{-2}$) and line-of-sight velocity dispersion
  ($\sigma=17$~\kms) that are consistent with the observed properties
  of extended \hi\ disks in the local universe.  Our analysis of the
  rotation curve also suggests that this $r\approx30$~kpc gaseous disk
  is warped with respect to the stellar disk.  In addition, we detect
  two weak \mgii\ absorbers in the halo with small velocity
  dispersions ($\sigma<10$~\kms).  While the exact geometry is
  unclear, one component is consistent with an extraplanar gas cloud
  near the disk-halo interface that is co-rotating with the disk, and
  the other is consistent with a tidal feature similar to the
  Magellanic Stream.  We can place lower limits on the cloud sizes
  ($l>0.4$~kpc) for these absorbers given the extended nature of the
  background source.  We discuss the implications of these results for
  models of the geometry and kinematics of gas in the circumgalactic
  medium.

\end{abstract}


\section{Introduction}

The interplay between inflows and outflows of gas around galaxies has
wide-ranging implications for galaxy evolution.  Of particular
interest are the roles of gas inflows and gas cooling for fueling the
growth of galaxies \citep[e.g.,][]{whi78, whi91, ker05, dek06, van11,
  fau11} and the need to suppress star formation with feedback from
massive stars and supermassive black holes to explain the observed
galaxy luminosity function and color bimodality
\citep[e.g.,][]{ben03,cro06,opp10,gab11,vog13,hop14}.  The accretion
of gas onto dark matter halos is well understood in the context of our
$\Lambda$CDM paradigm, but the kinematics and multi-phase structure of
gas within halos are not uniquely predicted
\citep[e.g.,][]{spr02,age07,kau09,ker09a,jou12,nel13}.  This is an
important problem, and its solution requires knowledge of both the
accretion and feedback processes that regulate the efficiency with
which halos convert their baryons into stars
\citep[$\Omega_*/\Omega_b\approx5\%$; e.g.,][]{bel03,mou13,pla13}.

Cool gas clouds ($T\sim10^4$~K) in galaxy halos are of particular
interest because they provide a source of fuel for subsequent star
formation \citep[e.g.,][]{ker09h,put12}.  They also play an important
role in the acquisition of angular momentum for galaxy disks
\citep[e.g.,][]{pic11,ste13}.  These clouds can originate from cool
inflows along filaments \citep[e.g.,][]{fum11,ste11b}, galactic winds
\citep[e.g.,][]{hec00,vei05}, accretion of gas previously ejected in a
wind \citep[e.g.,][]{bre80,for14}, thermal instabilities in a hot halo
\citep[e.g.,][]{mo96,mal04}, and tidal interactions
\citep[e.g.,][]{yun99,bes12}.

Observations of $\mgii~\lambda\lambda$2796,2803 absorption-line
systems are useful for characterizing the cool gas in galaxy halos
\citep[e.g.,][]{ber91,ste94,chu96}.  Studies based on large samples of
\mgii\ absorbers ($N\sim10^3$) have used continuum and line emission
\citep[e.g.,][]{zib07,mena11} and clustering
\citep[e.g.,][]{bou06,lun09,gau09} to establish a general trend that
stronger absorbers are more likely to be associated with star-forming
galaxies.  These trends with color or specific star-formation rate are
not as obvious in moderately sized samples ($N\sim10^2$) that
associate individual \mgii\ absorbers with individual galaxies
\citep[e.g.,][]{hchen10b,nie13}, but it is clear that the average
equivalent width and covering fraction are larger at smaller impact
parameters.  In particular, \mgii\ halos extend to
$\rho\approx100$~kpc for $L^*$ galaxies with a covering fraction of
$\approx50$\% for absorbers with
$W_r(2796)\geq0.5$~\AA\ \citep{hchen10a}, and the covering fraction at
$\rho<25$~kpc approaches unity \citep[e.g.,][]{kac13}.

Beyond these trends with impact parameter, there is strong evidence
that the equivalent width and covering fraction of \mgii\ absorbers
depends on the orientation of the background sightline with respect to
the disk of the foreground galaxy.  For example, when stacking
low-resolution spectra of $z_{\textnormal{\scriptsize{phot}}}>1$
galaxies behind $0.5<z_{\textnormal{\scriptsize{spec}}}<0.9$ galaxies,
\citet{bor11} found that edge-on disk galaxies at small impact
parameter ($\rho<40$~kpc) exhibit three times stronger absorption
along the minor axis than along the major axis.  This is consistent
with the signature of bipolar outflows driven perpendicular to the
plane of the disk, as observed in local starburst galaxies
\citep[e.g.,][]{bla88,sho98,oh02}.  It is also consistent with the
incidence of outflowing gas traced by \mgii\ absorption in
down-the-barrel sightlines towards star-forming galaxies, both for
stacked spectra \citep[e.g.,][]{wei09,rub10b} and for spectra of
individual galaxies
\citep[e.g.,][]{tre07,dia12,kor12,erb12,mar12,rub14}.  This empirical
evidence for winds perpendicular to the disk has motivated studies of
the spatial extent, mass, and kinematics of outflowing gas for
individual galaxies based on sightlines towards background quasars
\citep[e.g.,][]{tri11,bou12,gau12,lun12}.

In addition to a preference for absorption along the minor axis, there
is evidence for \mgii\ absorption associated with the plane of the
disk.  In particular, \citet{bou12} and \citet{kac12} argued that the
distribution of azimuthal angles for individual \mgii\ absorbers is
bimodal, with gas being found preferentially within
$\approx20^{\circ}$ of either the minor or major axis.  While it is
challenging to characterize the detailed three-dimensional geometry of
cool halo gas based on individual sightlines, there are clear examples
for which \mgii\ absorbers are co-rotating with the stellar disk
\citep[e.g.,][]{ste02,che05,kac10,kac11a,bou13}.  For example,
\citet{ste02} compared the kinematics of \mgii\ absorbers at
$\rho=10$--50~kpc to emission-line rotation curves at $r\leq10$~kpc
and found that the absorption-line velocities were often offset from
the systemic redshift in the sense of the galaxy's rotation.  In
detail, the full velocity ranges of \mgii\ absorption ($\Delta
v\approx100$--200~\kms) were broader than expectations for a single
thick disk, but the overall kinematics were clearly dominated by
rotation.

While absorption-line studies of halo gas have typically used
sightlines to background quasars, the advent of large redshift surveys
and sensitive spectroscopy with 8--10~m class telescopes has enabled
the use of background galaxies as light beacons.  This has been done
in both stacked \citep[e.g.,][]{ste10,bor11} and individual spectra
\citep[e.g.,][]{ade05,rub10a,ste10} at low spectral resolution.
Compared to background quasars, the extended light profiles of
background galaxies provide an opportunity to study absorption over a
larger solid angle through a given halo \citep[e.g.,][]{rub10a}, and
the number density of galaxies on the sky offers significantly more
sightlines to study halo gas in absorption at faint magnitudes
\citep[e.g.,][]{bar08}.  

However, it has not yet been possible to obtain spectroscopic data for
individual background galaxies that are of comparable quality (e.g.,
S/N, spectral resolution) to what is available for background quasars.
In particular, high-resolution spectroscopy is required to measure the
kinematics, column density, and covering fractions for individual
absorption components, and these measurements provide important
constraints for models of cool gas clouds in the circumgalactic
medium.

In this paper, we use a bright galaxy from the \citet{dia12} sample as
a background light source ($z_{bg}=0.712$) to probe halo gas
associated with a disk galaxy ($z_{fg}=0.413$) at impact parameter
$\rho=27$~kpc.  The brightness of the background galaxy ($g=19.6$)
offers the first opportunity to study halo gas at high spectral
resolution towards an extended background source.  We describe the
data used in this study in Section~2.  We present results on the
properties of the foreground disk galaxy and its circumgalactic gas in
Section~3 and 4.  We discuss the implications of these results in
Section 4, and we summarize our main conclusions in Section 5.
Throughout the paper, we adopt a cosmology with $h=0.7$,
$\Omega_M=0.3$, and $\Omega_{\Lambda}=0.7$.

\begin{figure*}[!t]
\begin{center}
\includegraphics[angle=0,scale=.85]{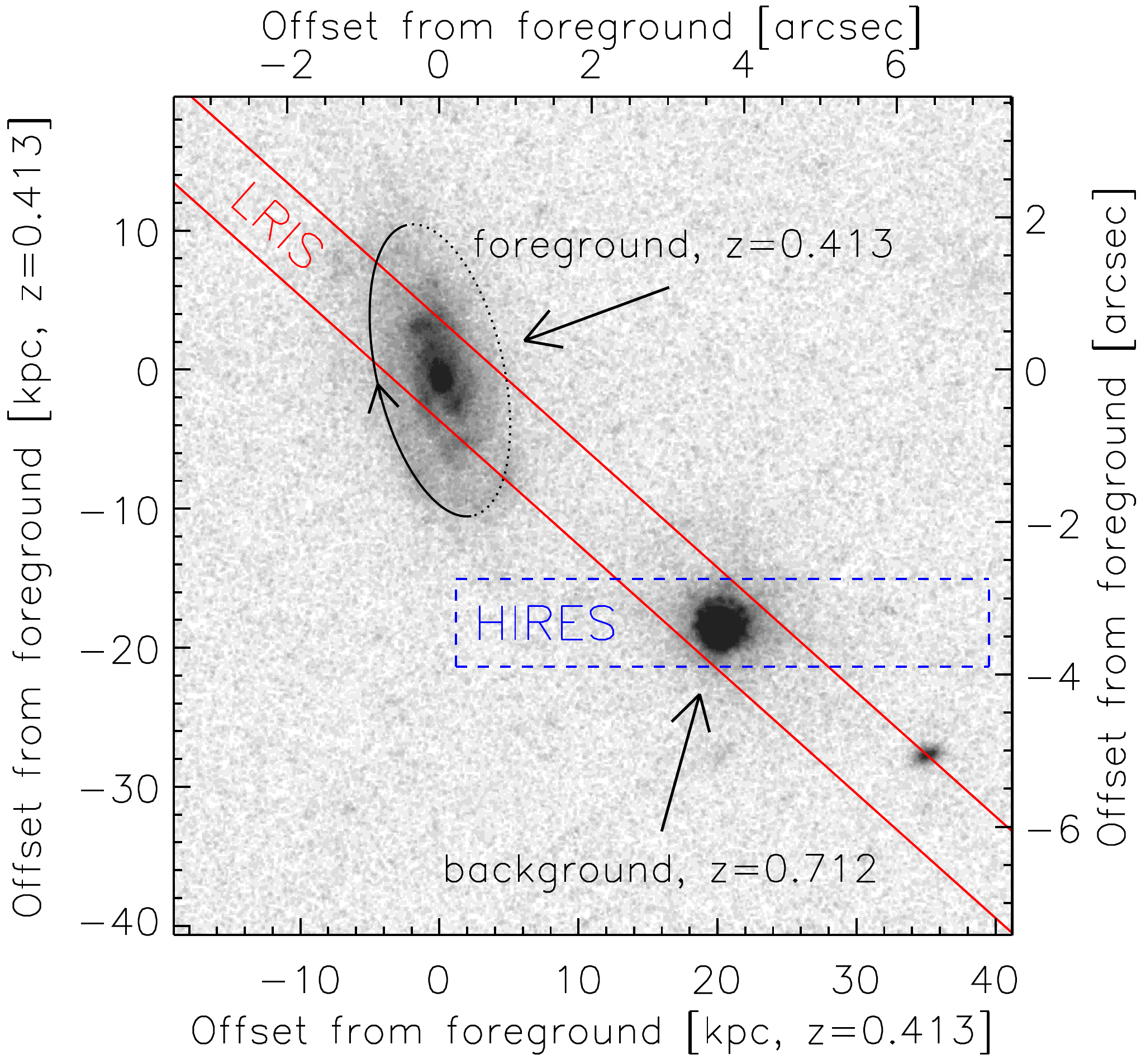}
\caption{HST WFC3/UVIS F814W image of the background and foreground
  galaxies described in this paper.  The positions and orientations of
  the Keck/LRIS (solid red lines) and Keck/HIRES (dashed blue
  rectangle) slits are shown.  The orientation of the foreground disk
  is shown by an ellipse, and the direction of rotation is indicated
  by the arrow on the near side of the disk.  The dotted portion of
  the ellipse corresponds to the far side of the disk, which is
  oriented into the page.}
\label{fig:hst_slits}
\end{center}
\end{figure*}

\begin{figure*}[!t]
\begin{center}
\includegraphics[angle=0,scale=.45]{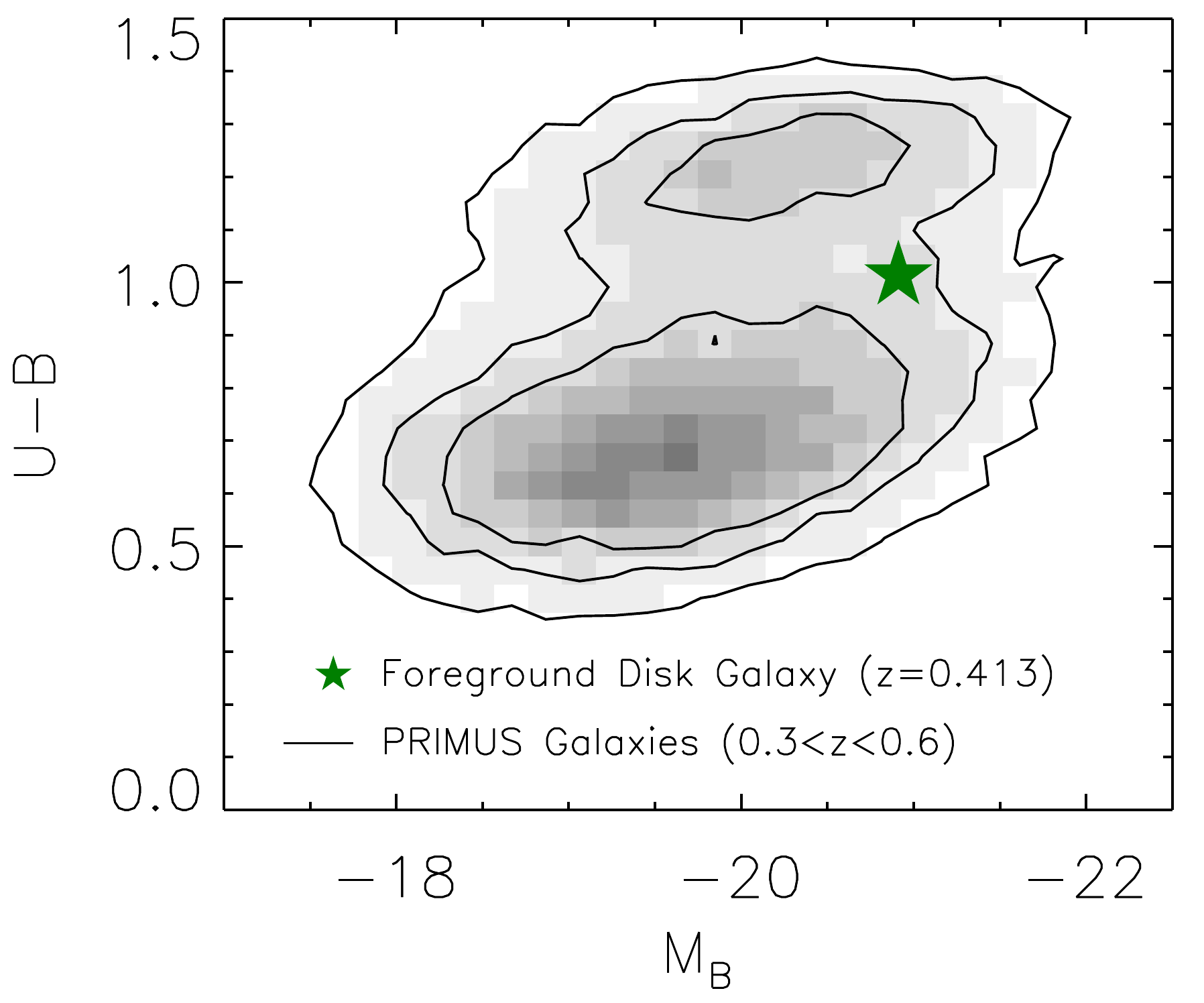}
\hspace{1cm}
\includegraphics[angle=0,scale=.45]{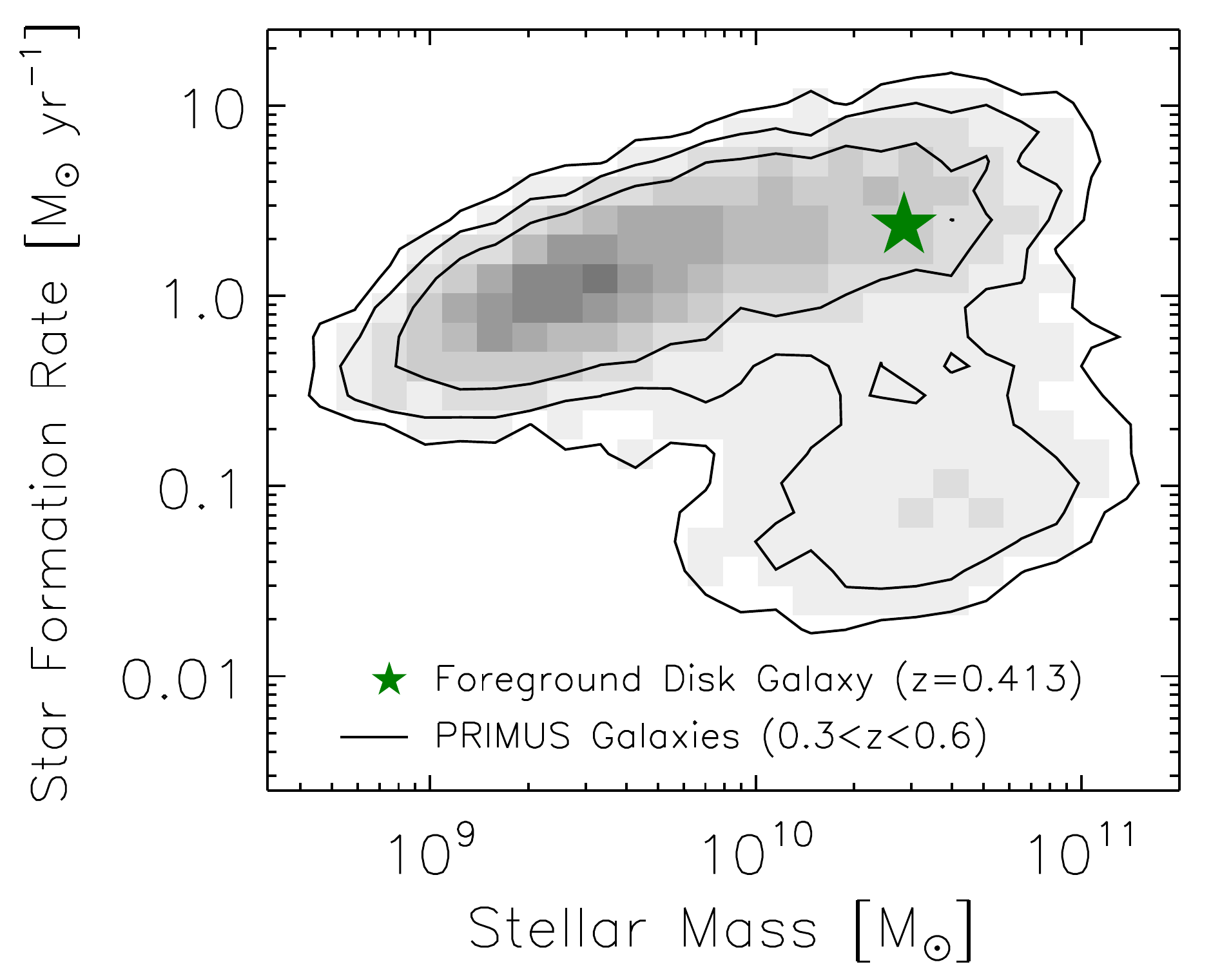}
\caption{{\it Left}: This color-magnitude diagram shows the location
  of the foreground galaxy with respect to a comparison sample of
  $\approx50,000$ galaxies at comparable redshifts from the PRIMUS
  survey.  The contours encompass 50\%, 75\%, and 90\% of the galaxies
  from the comparison sample.  The $U-B$ color of the foreground
  galaxy place it near the luminous end of the green valley.  {\it
    Right:} The star-formation rate vs stellar mass plane for the same
  set of galaxies, which illustrates the location of the foreground
  galaxy near the massive end of the star-forming sequence. }
\label{fig:fg_cmd}
\end{center}
\end{figure*}

\section{Data}

\subsection{HST/WFC3}

Observations with the Wide Field Camera 3 on the Hubble Space
Telescope were obtained on 2010 Dec 15 as part of program 12272.  As
described by \citet{dia12}, we obtained $4\times10$~min exposures in a
single orbit using the F814W filter on the UVIS channel (pixel size
0.04\arcsec), and we combined the dithered images to produce a science
mosaic with 0.02\arcsec\ pixels.  An $11\arcsec\times11\arcsec$
portion of the image including the foreground disk galaxy and the
compact background galaxy is shown in Figure~\ref{fig:hst_slits}.

\subsection{Keck/LRIS}

Observations with the Low Resolution Imaging Spectrometer
\citep[LRIS,][]{oke95} on the Keck~I telescope were obtained on 2011
Mar 7.  Using a 1\arcsec\ longslit and the D560 dichroic, we obtained
a 30-min exposure with the 400/3400 grism on the blue side
($R\approx600$) and $2\times15$~min exposures with the 400/8500
grating on the red side ($R\approx1000$).  The slit position angle was
set to -41.9~deg to include both the background (SDSS
J090523.59+575912.4) and foreground (SDSS J090524.08+575915.9)
galaxies (see the solid red lines in Figure~\ref{fig:hst_slits}).  The
data were processed with the XIDL
LowRedux\footnote{http://www.ucolick.org/\textnormal{\~}xavier/LowRedux/}
pipeline.

\subsection{Keck/HIRES}

Observations of the background galaxy with the High Resolution Echelle
Spectrometer \citep[HIRES,][]{vog94} on the Keck~I telescope were
obtained on 2009 Dec 11.  We obtained $2\times1$~hr exposures using
the HIRESb configuration with a $1.148\arcsec\times7.0\arcsec$ slit
(see the dashed blue rectangle in Figure~\ref{fig:hst_slits}), which
provides a spectral resolution of $R\approx37,000$ and a velocity
resolution of 8~\kms.  The data were processed with the XIDL
HIRedux\footnote{http://www.ucolick.org/\textnormal{\~}xavier/HIRedux/}
pipeline.  The spectrum of the background galaxy is fairly smooth near
the foreground absorption lines \citep{dia12}, and we fit the
continuum for each order using a low-order polynomial, excluding
regions of line absorption and emission.  The continuum-normalized
data from the two exposures and from overlapping regions between the
orders were combined using an inverse-variance weighted mean.

\subsection{Spitzer/IRAC}

Observations with the Infrared Array Camera \citep[IRAC,][]{faz04} on
the Spitzer Space Telescope were obtained on 2009 Nov 27 as part of
Program 60145.  We obtained $5\times30$~sec dithered exposures in the
3.6~$\mu$m and 4.5~$\mu$m bands.  We performed photometry on the
post--basic calibrated using the MOPEX point-source extraction
software \citep{mak05}.  Given the close spatial separation of the
background and foreground galaxies ($5.2\arcsec$) and the spatial
resolution of IRAC ($\textnormal{FWHM}=1.7\arcsec$), we used
point-source photometry to properly deblend the two galaxies.

\begin{figure*}[t]
\begin{center}
\includegraphics[angle=0,scale=.55]{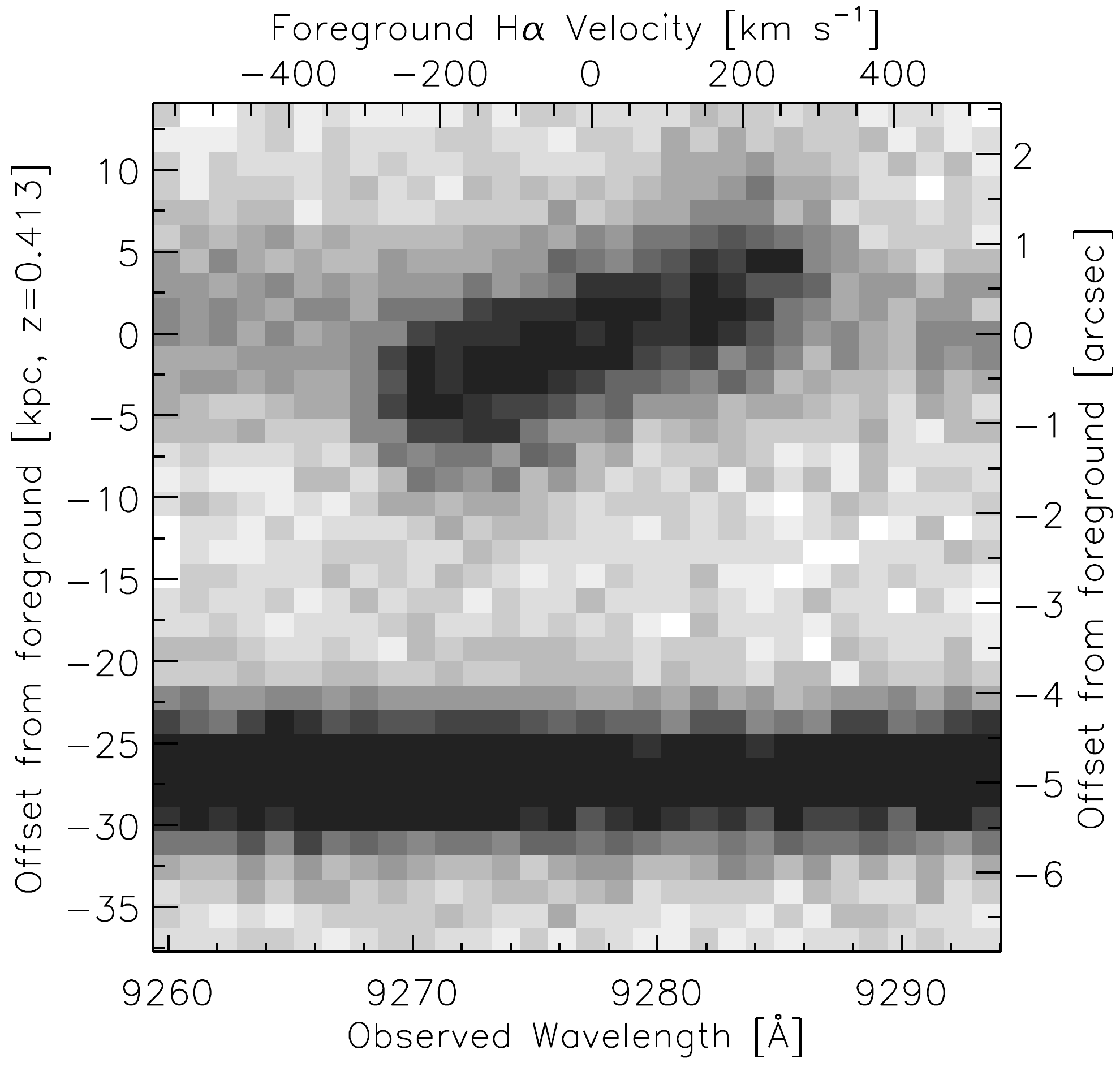}
\caption{The two-dimensional spectrum from Keck/LRIS near
  $\lobs=9276$~\AA, showing the H$\alpha$ emission from the foreground
  galaxy (top) and the bright continuum of the background galaxy
  (bottom).  The H$\alpha$ emission from the foreground galaxy is
  spatially extended and exhibits the signature of rotation.  The
  ionized gas in the disk of foreground galaxy is blueshifted by
  $v\approx-200$~\kms\ on the side closest to the background galaxy.}
\label{fig:ha_lris}
\end{center}
\end{figure*}

\section{Properties of the Foreground Disk Galaxy}

\subsection{Redshift, Rest-frame Color, Stellar Mass} 

We measure a systemic redshift of $z=0.4134\pm0.0001$ for the
foreground disk galaxy based on \oii~$\lambda3727$,
\oiii~$\lambda5007$, and H$\alpha$~$\lambda6563$ emission lines in the
Keck/LRIS spectrum.  We then use K-correct \citep{bla07} to estimate
rest-frame color ($U-B=1.01$) and absolute magnitude ($M_B=-20.91$)
based on photometry from the Sloan Digital Sky Survey
\citep[SDSS][]{aba09} and from Spitzer/IRAC.  We estimate the stellar
mass ($\mstar\approx10^{10.46}~\msun$) and star formation rate
($\textnormal{SFR}\approx2.4~\msun~\textnormal{yr}^{-1}$) from the
same photometry using iSEDfit, the spectral energy distribution
modeling code described by \citet{mou13}.  These values are derived
following the fiducial method from \citet{mou13}, which uses Flexible
Stellar Population Synthesis models \citep{con09,con10}, a
\citet{cha03} initial mass function from 0.1 to $100~\msun$, and the
time-dependent dust attenuation curve of \citet{cha00}.

We show the location of the foreground disk galaxy on a
color--magnitude diagram and a star formation rate--stellar mass
diagram in Figure~\ref{fig:fg_cmd}.  As a comparison sample, we
include $\approx50,000$ galaxies at $0.3<z<0.6$ from the Prism
Multi-object Survey\footnote{http://primus.ucsd.edu/}
\citep[PRIMUS,][]{coi11p,coo13,mou13}.  This figure illustrates that
the foreground disk galaxy resides on the luminous end of the green
valley \citep[e.g.,][]{men11} and the massive end of the star-forming
sequence \citep[e.g.,][]{noe07}.  The dust attenuation we estimate
from iSEDfit is $A_V\approx0.5$, which is consistent with a
star-forming galaxy viewed at moderate inclination.  The stellar mass
and luminosity of the foreground galaxy place it near $M^*$ for the
stellar mass function \citep{mou13} and $L^*$ for the luminosity
function at $z\approx0.4$ \citep{wilm06}.  In particular, the
foreground galaxy would be (1.2, 1.1, 0.8)$\times L^*$ as defined by
the $B$-band luminosity function for (blue, red, all) galaxies at
$z=0.5$ from \citet{wilm06}.

\subsection{Rotation curve and dynamical mass}\label{sec:dyn}

We show the two-dimensional Keck/LRIS spectrum in the region
surrounding the H$\alpha$ emission line for the foreground galaxy in
Figure~\ref{fig:ha_lris}.  The H$\alpha$ emission is spatially
extended and exhibits the signature of rotation with a line-of-sight
amplitude $v_{los}=178\pm10$~\kms.  We can convert this into a
circular rotation velocity ($v_{rot}$) by accounting for the disk
inclination ($i$) and the azimuthal angle of the slit relative to the
major axis of the disk ($\alpha$).  Following \citet{che05} and
\citet{che14}, this conversion can be expressed in terms of in terms
of $i$ and $\alpha$ for any location in the disk plane.
\begin{equation}\label{eqn:che}
\frac{v_{rot}}{v_{los}}=\frac{\sqrt{1+\sin^2{\alpha}
    \tan^2{i}}}{\cos{\alpha} \sin{i}}
\end{equation}

We fit an exponential profile to the HST data with GALFIT
\citep{pen02} to estimate the axis ratio ($b/a=0.39$) and position
angle of the disk ($\textnormal{PA}=15.8^{\circ}$).  This implies a
disk inclination $i=67^{\circ}$ and an azimuthal angle
$\alpha=33^{\circ}$ of the background sightline relative to the major
axis of the foreground disk.  Given the systematic uncertainties
associated with estimating projection parameters for spiral galaxies
\citep[e.g.,][]{bar03}, we adopt $\pm5^{\circ}$ uncertainties on the
inclination and azimuthal angles.

Before using equation~\ref{eqn:che} to estimate a circular velocity,
we need to account for the fact that each spatial bin of the LRIS
spectrum includes emission from regions that span a large range in
azimuthal angle.  For example, Figure~\ref{fig:hst_slits} shows that
the brightest regions of the galaxy that fall within the slit are
close to the major axis (i.e., they have small $\alpha$ angles).  To
account for this, we determine all the locations in the galaxy that
would fall in each 1.0'' wide $\times$ 0.2'' long spatial bin along
the LRIS slit.  We calculate the relevant $\alpha$ angle for all $500$
pixels in the bin and then determine the appropriate $v_{rot}/v_{los}$
ratio by computing a flux-weighted average, assuming that the
$H\alpha$ morphology is similar to the observed HST morphology at
$\lrest\approx5800$~\AA.  For the spatial bins that correspond to the
flat part of the observed rotation curve, we find that this ratio
converges to $v_{rot}/v_{los}=1.51\pm0.05$, which corresponds to a
circular rotation velocity $v_{rot}=269\pm18$~\kms.

The GALFIT profile fitting also yields an estimate of the half-light
radius, $r_e=5.7\pm0.3$~kpc.  We estimate the uncertainty in this
value from variations between fits that include Sersic profiles ($n$
is a free parameter) and exponential profiles ($n=1$), and fits that
include a second component to model emission from the central
kiloparsec of the galaxy.  For the two-component fits, the fainter
central component contributes $7\pm2\%$ of the total light and has
$r_e=0.4\pm0.1$~kpc.  The best-fit Sersic index for this central
component is $n=0.6\pm0.1$, which indicates that it does not exhibit
the characteristics of a classical bulge (i.e., it is not well fit by
an $n=4$ de Vaucouleurs profile).

We note that a circular velocity $v_c=269$~\kms\ at $r_e=5.7$~kpc
corresponds to a dynamical mass $M_{dyn}(r<r_e)\approx r_e v_c^2 / G =
9.6\times 10^{10}~\msun$, which is 0.5~dex larger than our estimate of
the stellar mass above.  This offset is somewhat larger than the
0.3~dex systematic uncertainty in the stellar mass estimate, so this
suggests that the dynamical mass has comparable contributions from
baryons and dark matter within the effective radius.

\begin{figure*}[!t]
\begin{center}
\includegraphics[angle=0,scale=0.85]{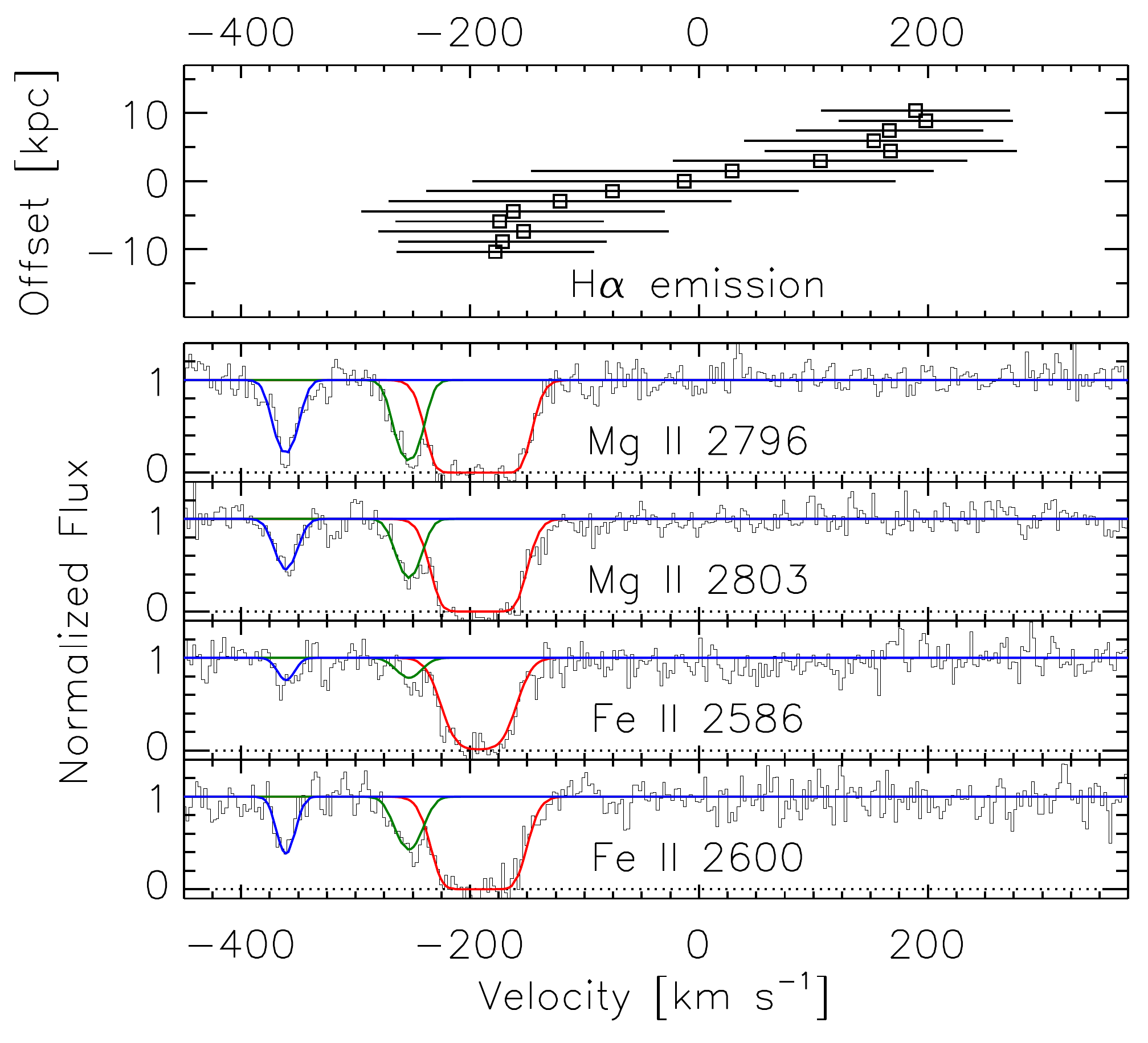}
\caption{Continuum-normalized spectra from Keck/HIRES showing
  absorption-line profiles for four transitions associated with the
  foreground galaxy (\mgii~2796, 2803; \feii~2586, 2600).  The solid
  red, green, and blue curves in each panel show the best-fit Voigt
  profiles for three components at $v=-195$~\kms, $v=-256$~\kms, and
  $v=-363$~\kms\ relative to the systemic redshift of the foreground
  galaxy ($z=0.4134$).  The kinematics of the strongest component at
  $v=-195$~\kms\ suggest that it is consistent with an extension of a
  flat rotation curve to $r\approx30$~kpc.}
\label{fig:fg_hires}
\end{center}
\end{figure*}

\section{Properties of Gas in the Circumgalactic Medium}

As discussed in the Introduction, the serendipitous superposition of
the background and foreground galaxies on the plane of the sky
presents an opportunity to study gas in the circumgalactic medium
(CGM) of the foreground galaxy at impact parameter $\rho=27$~kpc.  In
particular, we are able to measure the column density and kinematics
of gas in absorption along the line of sight to the background galaxy
at high spectral resolution ($\textnormal{FWHM}\approx8~$\kms).  In
this section, we present these measurements, and we describe the
kinematic modeling that is necessary to interpret the observed
line-of-sight velocities in terms of a rotating disk, a warped disk,
or an outflowing wind.

\subsection{Observed column densities and kinematics}\label{sec:nh}

We show the continuum-normalized Keck/HIRES data for the
\mgii~$\lambda\lambda$2796,2803 and \feii~$\lambda\lambda$2586,2600
absorption lines in Figure~\ref{fig:fg_hires}.  We also use spectral
coverage of three other \feii\ lines (2344, 2374, 2382) in our
analysis.  We detect an optically thick absorption component centered
near $v\approx-200$~\kms, along with two weaker components near
$v\approx-260$~\kms\ and $v\approx-360$~\kms.  We use
VPFIT\footnote{http://www.ast.cam.ac.uk/\textnormal{~}rfc/vpfit.html}
to perform a joint fit to all three components for both ions to
measure precise column densities and kinematics (see
Figure~\ref{fig:fg_hires} and Table~\ref{tab:foreground}).  In
particular, we require that the redshifts (i.e., centroid velocities)
for the three velocity component be the same for both \mgii\ and
\feii\, but we allow the $b$ values and column densities to be fit
independently for each ion.

Using reasonable assumptions about gas-phase abundances and ionization,
we can estimate the hydrogen column density and \mgfe\ for each
velocity component. For our fiducial estimates, we assume no dust
depletion and solar metallicity, and we assume that \mgii\ and
\feii\ are the dominant tracers of \mg\ and \fe, respectively.
Relaxing any of these assumptions would increase our estimates of
$N_H$ (i.e., dust depletion or lower metallicity would reduce the
gas-phase abundance of \mg\ and \fe\ relative to hydrogen, and other
ionization phases could contribute additional \mg\ and \fe).  Based on
the \feii\ column densities, these fiducial assumptions imply hydrogen
column densities $N_H\geq1.2\times10^{19}$~cm$^{-2}$ for the strongest
component and $N_H\geq3$--$4\times10^{17}$~cm$^{-2}$ for the two
weaker components.  We note that the estimated dust depletion for
\fe\ in the warm ionized medium is $-0.8$~dex
\citep[e.g.,][]{jen09,dra11} and Lyman Limit Systems are typically
associated with sub-solar metallicity \citep[e.g.,][]{leh13}, so these
$N_H$ values could be larger by an order of magnitude.

Based on the \mgii\ and \feii\ column densities, all three velocity
components have $\mgfe\approx-0.2$.  The \mgii\ lines for the
strongest component are saturated, which implies that there could be
contributions from unresolved velocity subcomponents.  That said, the
\feii\ lines span a much wider range in optical depth, and the $b$
values we measure for the \mgii\ and \feii\ components are nearly
identical.  This suggests that our column density measurements for
both ions are robust, even for this strongest component.  We note that
correcting for dust depletion would further decrease $\mgfe$ because
\mg\ is less depleted than \fe\ \citep[e.g.,][]{jen09,dra11}.  The
small \mgfe\ ratio for this absorbing gas suggests that it has been
significantly enriched in \fe\ by previous generations of type 1a
supernovae \citep[e.g.,][]{rig02}.

\subsection{Kinematics of gas in the disk plane}\label{sec:disk}

If the kinematics of these absorbers were dominated by disk rotation
in the plane defined by the stellar emission, we could convert the
observed line-of-sight velocities ($v_{los}$) into deprojected
rotation velocities ($v_{rot}$).  This depends on the sine of the disk
inclination angle ($i$) and the cosine of the azimuthal angle in the
plane of the disk ($\phi$).
\begin{equation}\label{eqn:vrot}
\frac{v_{rot}}{v_{los}}=\frac{1}{\sin{i}\cos{\phi}}
\end{equation}
Following \citet{ste02}, the angle $\phi$ can be expressed in terms of
the impact parameter measured along the major axis ($p$) and the
distance along the minor axis where the line of sight intersects the
disk plane ($y_{0}$).
\begin{equation}\label{eqn:phi}
cos{\phi}=\frac{p}{\sqrt{p^2 + y_{0}^2}}=\frac{1}{\sqrt{1+(y_0/p)^2}} 
\end{equation}
These values can be determined from the impact parameter ($\rho$) and
the azimuthal angle in the plane of the sky ($\alpha$).
\begin{equation}\label{eqn:p}
p=\rho \cos{\alpha} ~ ; ~ y_0 = \frac{\rho \sin{\alpha}}{\cos{i}}
\end{equation}
We note that substituting the values from equations~\ref{eqn:phi} and
\ref{eqn:p} into equation~\ref{eqn:vrot} yields the expression in
equation~\ref{eqn:che}.

Based on the above, we find $v_{rot}/v_{los}=2.1$, $p=22$~kpc,
$y_{0}=37$~kpc, $r=43$~kpc, and $\phi=59^{\circ}$.  We mark this
location with a red circle in Figure~\ref{fig:on}, which shows the
projection of the disk on the plane of the sky along with $y-z$
(edge-on) and $x-y$ (face-on) views of the disk plane.  The velocity
ratio implies deprojected rotation velocities $v_{rot}=[-410, -540,
  -760]$~\kms\ for the three absorption components.  All of these
values would be inconsistent with the Tully-Fisher relation
\citep[e.g.,][]{tul77,mcg00} and the observed rotational velocity of
the stellar disk (Section~\ref{sec:dyn}), which suggests that the
absorbers are either not located in the disk plane defined by the
stellar disk, or their kinematics are not dominated by rotation.

\subsection{Kinematics of gas in a warped disk}\label{sec:warp}

Our line of sight would also intersect clouds located above and below
the disk plane, and if the kinematics were dominated by rotation, the
$v_{rot}/v_{los}$ ratio would also be determined by
equation~\ref{eqn:vrot}.  The relevant $\phi$ angle would depend on
the $y$ coordinate along the minor axis, which can be calculated for
any point along the line-of-sight vector $\mathbf{L}$ by considering
its projection along the $y$ and $z$ planes.

\begin{equation}\label{eqn:los}
x = p ~ ; ~ y = y_{0} - |\mathbf{L}| \sin{i} ~ ; ~ z = z_{0} + |\mathbf{L}| \cos{i}
\end{equation}

Here the $x$ coordinate is measured along the major axis in the disk
plane, the $y$ coordinate is measured along the minor axis in the disk
plane, and the $z$ coordinate is measured along the vertical axis
above the disk plane.  For example, the reference position
$y_0=37$~kpc corresponds to $z_0=0$ (i.e., crossing the disk plane),
and a location $| \mathbf{L} |=40$~kpc along the line of sight from
this reference position would have $y=0$ and $z=16$~kpc (i.e.,
crossing the major axis of the disk).  The $y$ coordinate then
determines $v_{rot}/v_{los}$ for every point along the line of sight
based on equations~\ref{eqn:vrot} and \ref{eqn:phi}.

Therefore, we can determine the location that would yield an observed
line-of-sight velocity for a given rotation velocity.  If we use the
rotation velocity of the stellar disk ($v_{rot}=269$~\kms), the
strongest absorber could be at $| \mathbf{L} |=21$~kpc ($z=8$~kpc,
$r=30$~kpc, $v_{rot}/v_{los}=1.4$), which is marked by a red star in
Figure~\ref{fig:on}.  If this location were associated with a warped
disk, the relevant warp angle would be $\tan^{-1}{z/r}=16^{\circ}$.

\subsection{Kinematics associated with outflowing gas}\label{sec:outflow}

If the kinematics of the absorbers were dominated by an outflowing
wind, we could convert the observed line-of-sight velocities into
three-dimensional wind velocities.  The ratio of the wind velocity
($v_{w})$ to the line-of-sight velocity ($v_{los}$) can be expressed
in terms of the angle ($i_{w}$) between the wind vector and the line
of sight.
\begin{equation}
\frac{v_{w}}{v_{los}}=\frac{1}{\cos{i_{w}}}
\end{equation}
For a wind vector that is orthogonal to the disk plane (i.e., along
the $z$ axis) the angle $i_w$ would be equal to the disk inclination
angle ($i$).  More generally, the angle $i_w$ can be expressed in
terms of the length of the wind vector ($r_{w}$) and its projection on
the plane of the sky (i.e., the impact parameter, $\rho$).
\begin{equation}
\sin{i_w} = \frac{\rho}{r_w}
\end{equation}
Following \citet{gau12}, $r_w$ can be expressed in terms of the $z$
height above the disk plane and the half opening angle of the wind
with respect to the $z$ axis ($\theta_w$).  
\begin{equation}\label{eqn:rw}
r_w = \frac{z}{\cos{\theta_w}}
\end{equation}
Thus for any location along the line of sight
(equation~\ref{eqn:los}), we can calculate the galactocentric radius
($r_w$), the opening angle of a wind vector that intersects the line
of sight ($\theta_w$), and the associated $v_w/v_{los}$ ratio.

\begin{figure*}[t]
\begin{center}
\includegraphics[angle=0,scale=.45]{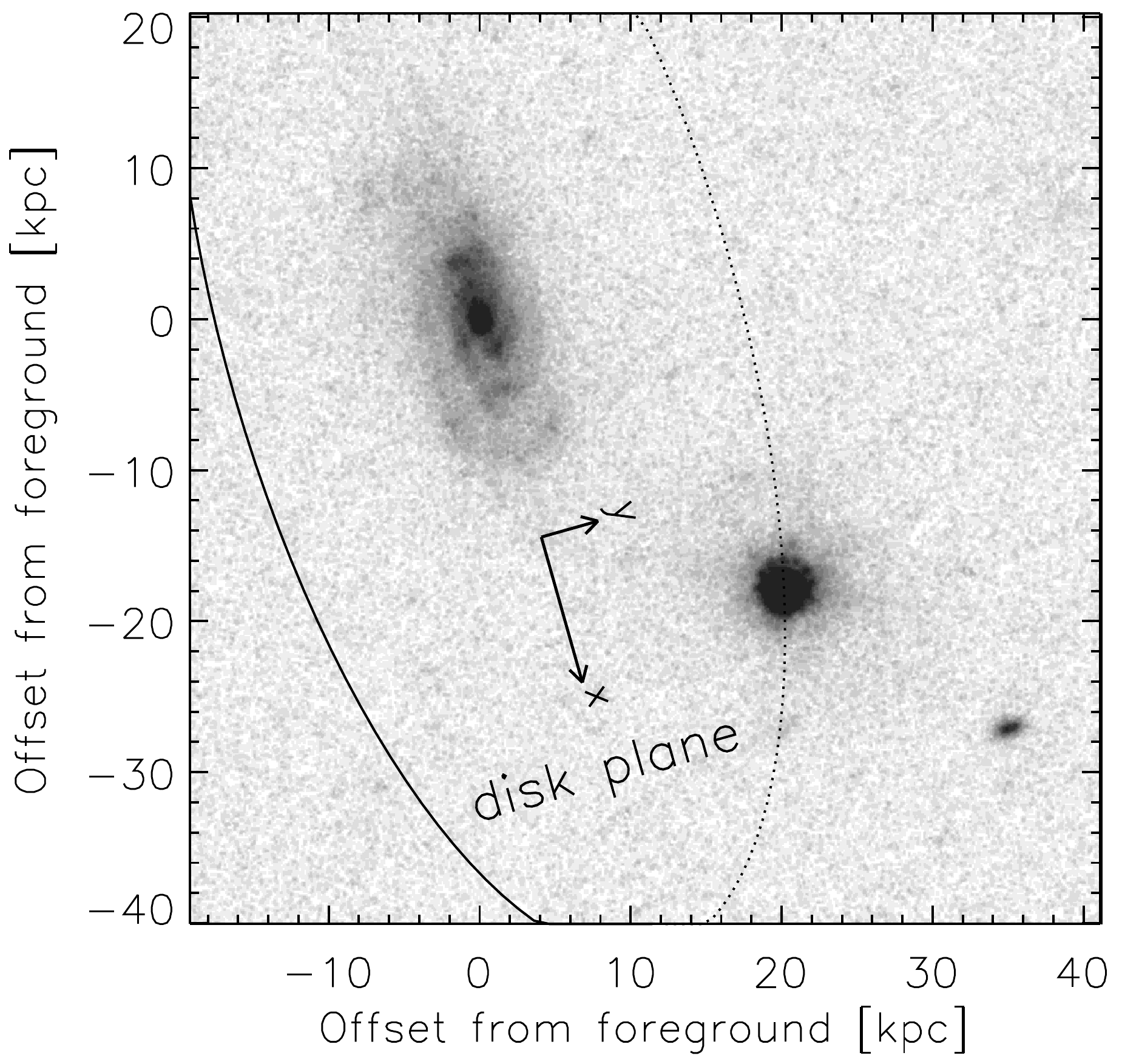}
\includegraphics[angle=0,scale=.45]{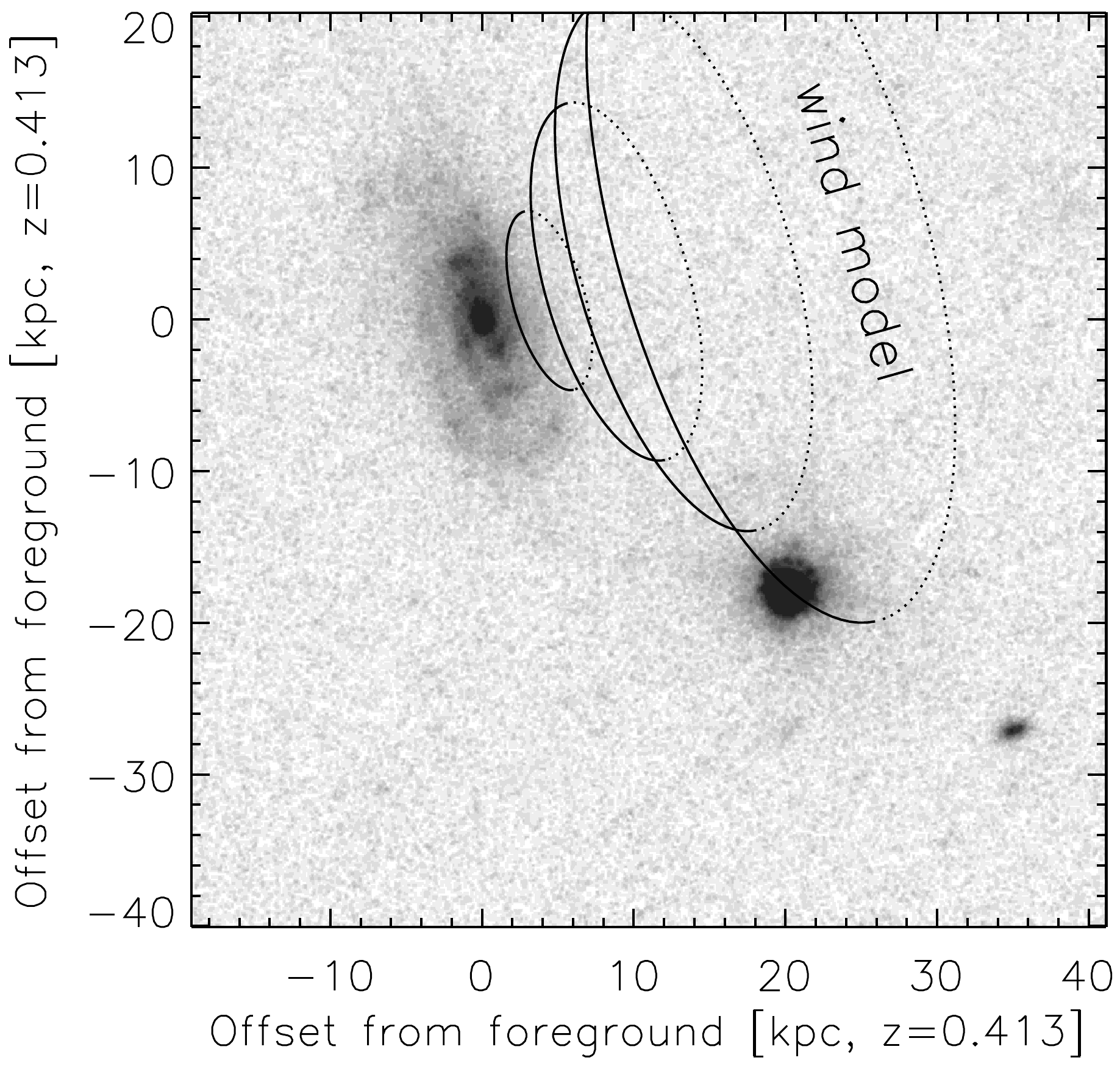}
\end{center}
\begin{center}
\includegraphics[angle=0,scale=.45]{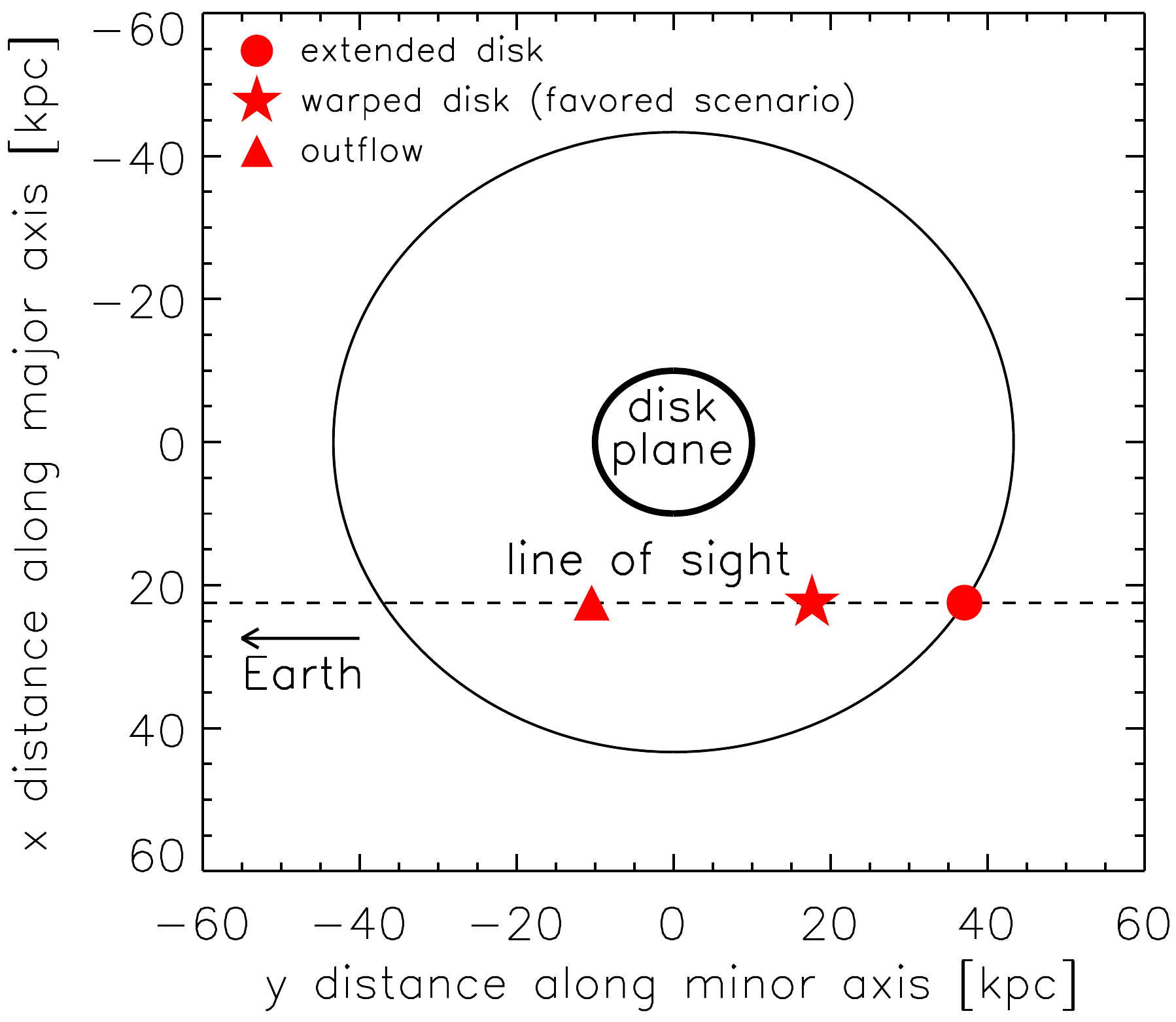}
\includegraphics[angle=0,scale=.45]{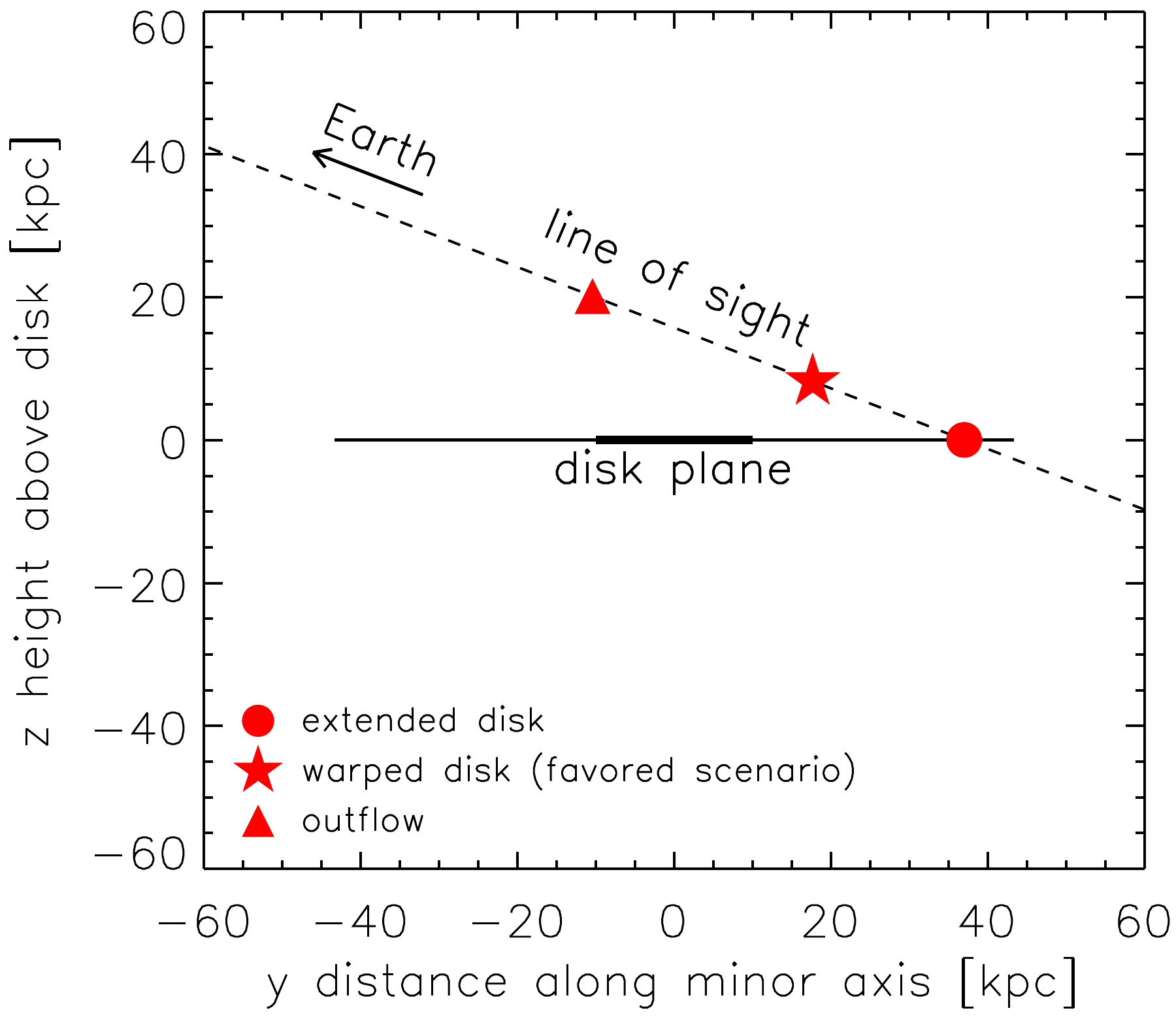}
\caption{{\it Top Left:} Visualization of the disk plane projected on
  the plane of the sky.  The solid and dotted lines show a circular
  $r=43$~kpc disk in the same plane as the stellar disk.  The $x$ and
  $y$ arrows indicate a coordinate system with $x$ values measured
  along the major axis and $y$ values measured along the minor axis.
  {\it Top Right:} Visualization of a wind model with opening angle
  $\theta_w=51^{\circ}$ projected on the plane of the sky.  The solid
  and dotted lines correspond to locations in the outflow cone with
  $z$ heights of above the disk plane of 5~kpc, 10~kpc, 15~kpc, and
  21~kpc.  In this model the line of sight would intersect the edge of
  the outflow cone at $z=21$~kpc.  {\it Bottom Left:} A face-on view
  of the disk plane with an $r=10$~kpc black circle that represents
  the stellar disk, an $r=43$~kpc black circle that represents an
  extended disk, and a dashed line that represents the line of sight.
  The red points correspond to three potential scenarios that could
  explain the strongest absorption component at $v=-195$~\kms.  The
  red circle corresponds to an extended disk in the same plane as the
  stellar disk (Section~\ref{sec:disk}).  The red star corresponds to
  a warped disk that is consistent with a flat rotation curve
  (Section~\ref{sec:warp}).  The red triangle corresponds to the edge
  of an outflow cone that could intersect the line of sight
  (Section~\ref{sec:outflow}).  {\it Bottom Right:} An edge-on view of
  the disk plane with the same components as the left panel.  The
  $r=10$~kpc is drawn somewhat thicker than the $r=43$~kpc disk for
  visualization purposes.  The $z$ height is defined relative to the
  plane defined by the stellar disk.}
\label{fig:on}
\end{center}
\end{figure*}

We can therefore determine the minimum opening angle required to
intersect the line of sight and the maximum opening angle beyond which
one would expect a broader range of line-of-sight velocities than is
observed.  The minimum opening angle $\theta_{w}=51^{\circ}$
corresponds to $|\mathbf{L}|=55$~kpc ($z=21$~kpc, $v_w/v_{los}=1.6$),
and we mark this location with a red triangle in Figure~\ref{fig:on}.
For an outflow with an opening angle $\theta_{w}>51^{\circ}$, the line
of sight would enter (exit) the outflow cone at some lower (higher)
$z$ height for which a smaller (larger) fraction of its velocity would
be along the line of sight \citep[e.g.,][]{gau12}.  For example, an
outflow with $\theta_{w}=52^{\circ}$ would span a range
$|\mathbf{L}|=46$--68~kpc ($z=18$--26~kpc) and exhibit $v_w/v_{los}$
values that a vary by a factor of 1.9, which matches to the range of
absorption-line velocities we observe.  Therefore, in the context of a
conical outflow model, larger opening angles $\theta_{w}>52^{\circ}$
can be ruled out because they would be associated with an even broader
range of line-of-sight velocities than we observe.  We note that an
outflow model with $\theta_w>60^{\circ}$ would also be associated with
redshifted components, and a model with $\theta_{w}>67^{\circ}$ would
additionally produce a blueshifted ``down-the-barrel'' component in
self-absorption towards the foreground galaxy, neither of which is
observed.

\section{Discussion}

We have presented results on a Milky-Way-like disk galaxy at $z=0.4$,
including the column density and kinematics of its circumgalactic gas
as traced by \mgii\ and \feii\ absorption lines at impact parameter
$\rho=27$~kpc.  The CGM absorption lines are blueshifted in the sense
of the galaxy's rotation, and here we evaluate the hypothesis that
this gas is associated with an extended, warped gaseous disk.  We also
consider whether some components of this absorption could be
associated extraplanar halo gas.  This interpretation is relevant
because an extended disk may trace recent accretion of gas from the
intergalactic medium \citep[e.g.,][]{ost89,ste13}, and extraplanar
halo gas may trace outflows or gas recycling in a galactic fountain
\citep[e.g.,][]{sha76,bre80,for14}.  We also discuss the size
constraints that can be placed on the absorbing clouds due to the
extended size of the background source.  We present this discussion in
the context of previous observations of gas around galaxies as traced
by emission and absorption, and we compare to expectations from
simulations of gas in the circumgalactic medium.

\subsection{Evidence for an extended, warped disk}\label{sec:diswarp}

In the local universe, extended \hi\ disks have been observed in
emission around disk galaxies with spatial extents
$r_{\scriptsize{\hi}}=10$--50~kpc for mass surface densities
$\Sigma_{\scriptsize{\hi}}=1$~\msun~pc$^{-2}$
($N_H\approx10^{20}$~cm$^{-2}$); this spatial extent is a function of
\hi\ mass such that galaxies with
$M_{\scriptsize{\hi}}>10^{10}$~\msun\ generally have
$r_{\scriptsize{\hi}}>30$~kpc \citep[e.g.,][]{bro97}.  Most of these
\hi\ disks exhibit warps beyond the optical extent of the disk
\citep[e.g.,][]{san76,bos91,san08}, and the angles of these warps with
respect to the stellar disk can be as large as
$20^{\circ}$--$30^{\circ}$ \citep[e.g.,][]{gar02}.  In terms of
kinematics, these warped disks are also dynamically thin, with typical
velocity dispersions $\approx10$~\kms \citep[e.g.,][]{bos91,bot96}.
There are several explanations for the physical origin of warps,
including interactions with nearby dwarf galaxies
\citep[e.g.,][]{sha98} and cosmic infall of gas with misaligned
angular momentum \citep[e.g.,][]{ost89,jia99,she06}.

In the context of this paper, an extended disk provides a
straightforward explanation for the optically thick absorption
component at $v_{los}=-195$~\kms.  Given the incidence of extended
disks around massive, star-forming galaxies in the local universe, it
is reasonable to expect that our line of sight would intersect an
extended disk.  Galaxies at $z=0.4$ also tend to have higher gas
fractions than galaxies at $z=0$ \citep[e.g.,][]{rao06,com13}, which
provides additional support for this interpretation.  Furthermore, the
estimated \hi\ column density is consistent with the expected surface
density of an extended disk, and the kinematics agree with a flat
rotation curve ($v_{rot}\approx270$~\kms, Section~\ref{sec:warp}).
The velocity dispersion $\sigma=b/\sqrt{2}=17$~\kms\ is consistent
with the vertical kinematics of a relatively thin disk.

\subsection{Constraints on an outflowing wind}\label{sec:disoutflow}

Galactic winds are known to be common among star-forming galaxies at
$z\sim0.5$ \citep{mar12,rub14}, so it is worthwhile to consider
whether the absorption we observe could be associated with an
outflowing wind.  Estimates of the half-opening angle for winds range
from $\theta_w\approx30^{\circ}$ for nearby galaxies like M82 and
NGC~253 \citep{hec90,wes11} to
$\theta_w\approx40^{\circ}$--60$^{\circ}$ for larger samples of
star-forming galaxies at $z=0.1$--1
\citep{ychen10,bor11,kac12,mar12,rub14}.  In addition, \citet{mar12}
find that the opening angle may depend on velocity, such that higher
velocity outflows tend to be more collimated, while \citet{rub14} find
evidence that galaxies with larger star-formation rate surface
densities ($\sigmasfr$) tend to have larger wind opening angles.

As discussed in Section~\ref{sec:outflow}, only a small range of wind
opening angles $51^{\circ}<\theta_w<52^{\circ}$ could intersect our
light of sight without producing a broader range of line-of-sight
velocities than is observed.  Furthermore, this would require a very
fortunate geometry (i.e., the background source would have to be
almost exactly at the edge of the outflow cone), and it would also
require a large opening angle for a $|v_w|>300$~\kms\ outflow.  In
particular, the minimum opening angle $\theta_w=51^{\circ}$ would
require an outflow velocity $v_w\approx-320$~\kms
($v_w\approx-580$~\kms) to explain the lowest (highest) velocity
component we observe.  Furthermore, the maximum opening angle
$\theta_w=52^{\circ}$ would require $v_w\approx-460$~\kms\ to
reproduce both the low-velocity and high-velocity components.  While
there is evidence that galaxies with high $\sigmasfr$ values can
produce large-scale, high-velocity outflows
\citep[e.g.,][]{hec11,dia12,law12,sel14}, the foreground galaxy has
$\sigmasfr=SFR/(2\pi r_e^2)=0.01$~\msun~yr$^{-1}$.  This is an order
of magnitude below the canonical threshold for driving winds
\citep[$\sigmasfr\geq0.1$~\msun~yr$^{-1}$;][]{hec02} and is
approximately equal to the smallest $\sigmasfr$ value among $z\sim0.5$
galaxies that have been targeted in absorption-line studies of
galactic winds \citep[e.g.,][]{kor12,rub14}.  While it is clear that
large-scale galactic winds are capable of producing strong
\mgii\ absorption over a wide range of azimuthal angles around
star-forming galaxies with high $\sigmasfr$ values
\citep[e.g.,][]{lun12,rub14}, it is unlikely that the absorption lines
we observe around this low-$\sigmasfr$ galaxy are associated with an
outflowing wind.

\begin{figure*}[!t]
\begin{center}
\includegraphics[angle=0,scale=.45]{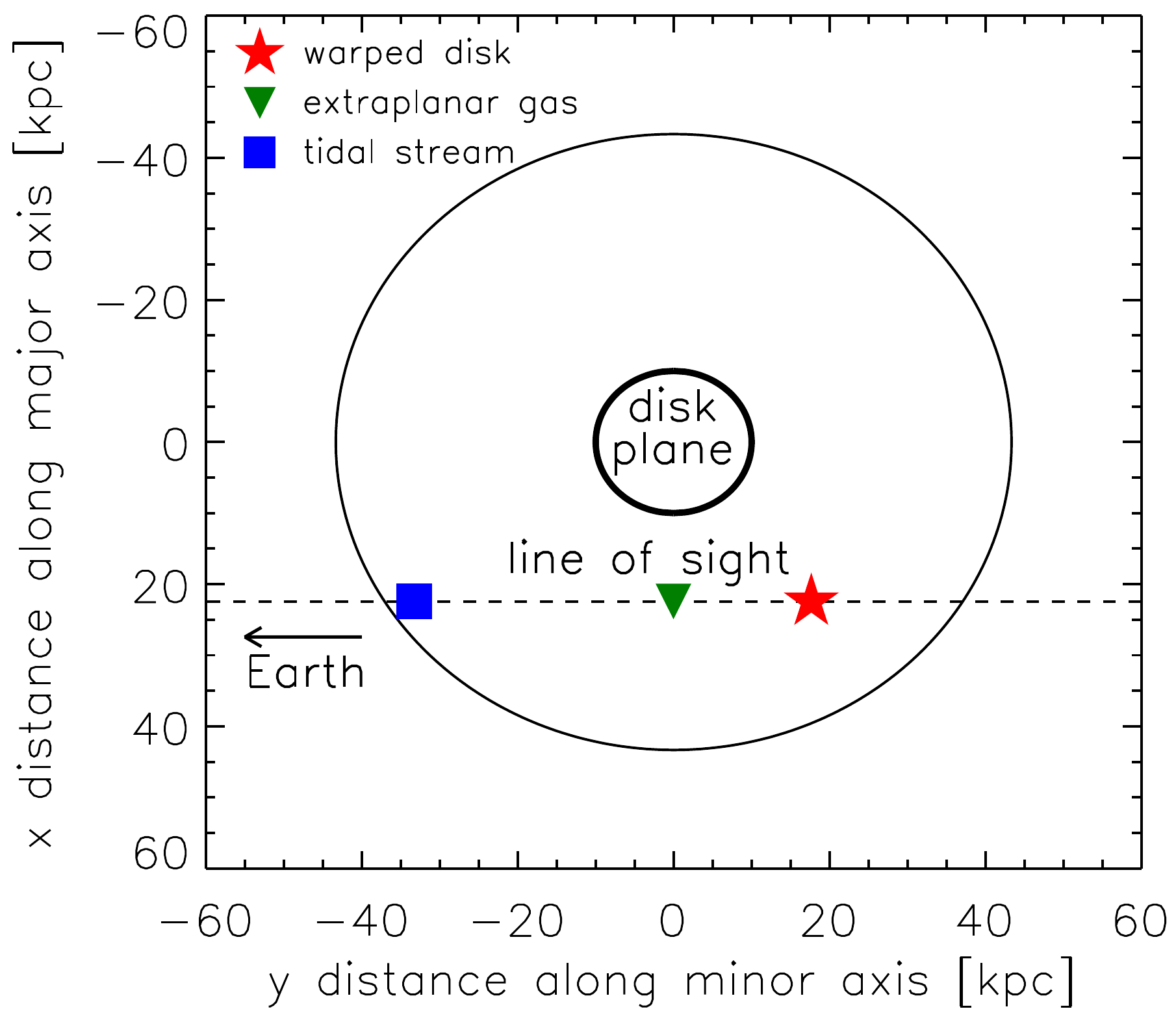}
\includegraphics[angle=0,scale=.45]{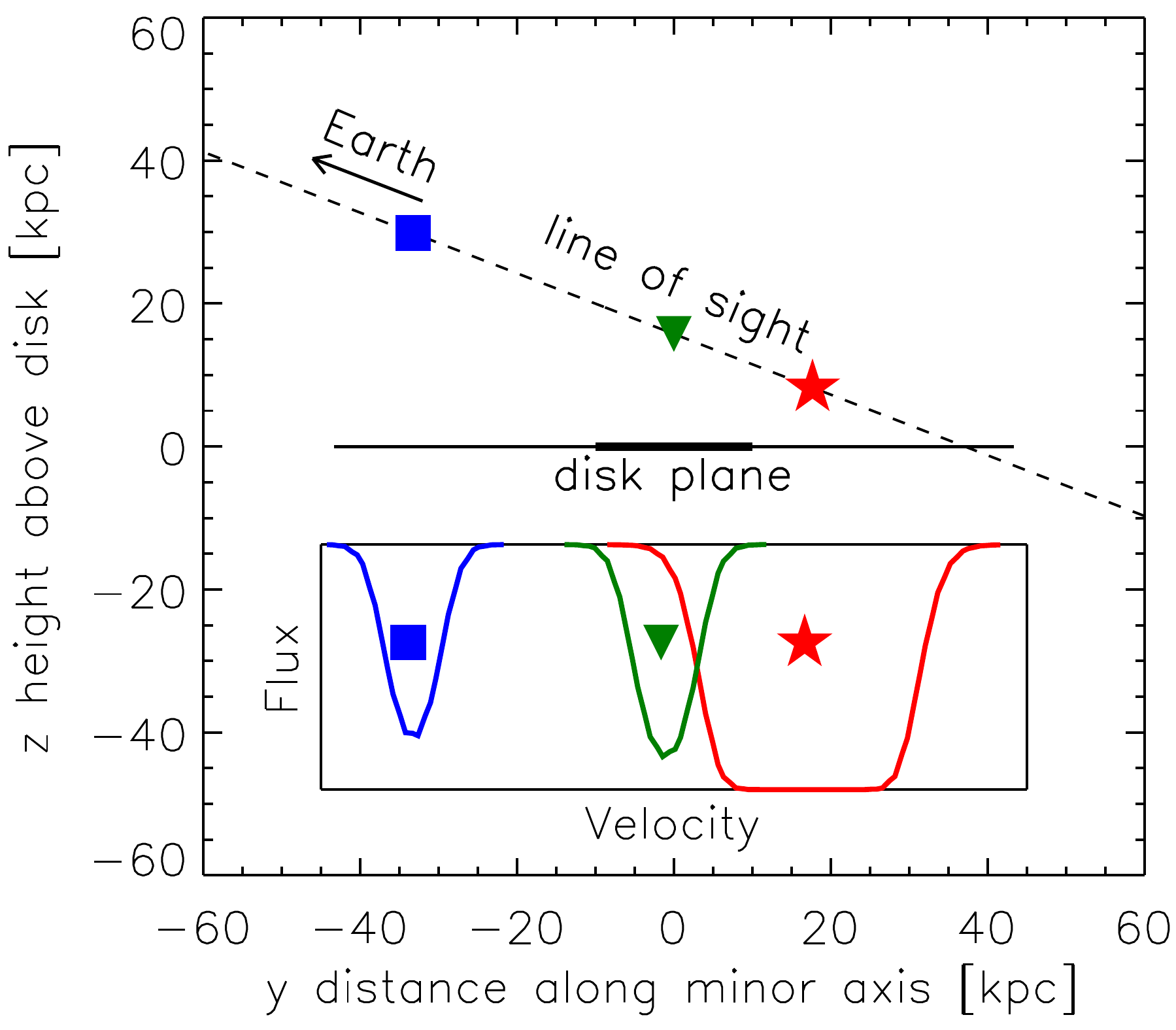}
\caption{Similar to Figure~\ref{fig:on}, the left and right panels
  show face-on and edge-on views of the disk plane.  The colored
  points correspond to physical scenarios that could explain each of
  the observed velocity components.  The red star corresponds to a
  warped disk, which provides the best explanation for the strongest
  component at $v=-195$~\kms\ (Section~\ref{sec:diswarp}).  The green
  triangle corresponds to co-rotating gas near the disk-halo interface
  that could explain the component at
  $v=-256$~\kms\ (Section~\ref{sec:disweak}).  The blue square
  corresponds to the potential location of a tidal stream that could
  explain the component at $v=-363$~\kms\ (Section~\ref{sec:disweak}).
  The inset in the right panel illustrates these three absorption
  components in velocity space.}
\label{fig:dis}
\end{center}
\end{figure*}

\subsection{Galactic fountains and recycled accretion}\label{sec:fountain}

The fact that we are tracing circumgalactic gas via \mgii\ and
\feii\ absorption lines indicates that this gas must have been
enriched by previous generations of supernovae (i.e., it is not
pristine, metal-free gas).  It is therefore worth considering models
of a galactic fountain, which describes the kinematics of
supernova-driven gas flows that eventually return to the disk
\citep[e.g.,][]{bre80,col02,fra06}, and models of recycled accretion
in cosmological simulations, which suggest that gas infall at $z<1$ is
dominated by material that was previously ejected in a wind
\citep[e.g.,][]{opp10,for14}.

The \citet{bre80} galactic fountain model describes how hot gas moving
vertically upward and radially outward would condense via thermal
instabilities into clouds that subsequently fall back towards the
disk.  If no external torques were applied to the gas, the orbital
angular momentum would be conserved such that gas clouds would follow
a ballistic orbit before returning to the disk near the radius where
the gas originated.  In the context of this paper, such a model would
not be able to explain the large specific angular momentum we observe
at $r\approx30$~kpc.  In other words, the rotational velocity of
supernova-heated gas originating from $r\leq5$~kpc would decrease by a
factor of six as it traveled radially outwards (e.g., decreasing from
$v_{rot}=270$~\kms\ at $r=5$~kpc to $v_{rot}=45$~\kms\ at $r=30$~kpc).
In contrast, the rotational velocities we infer at $r\approx30$~kpc
are comparable to the rotational velocities at $r\approx5$~kpc.  Thus
it is clear that a galactic fountain model in which the angular
momentum of ejected gas is conserved would be inconsistent with our
observations.

While the expectations for gas kinematics based on purely ballistic
orbits are well understood \citep[e.g.,][]{col02,fra06}, the angular
momentum of gas in a galactic fountain can also be modified by
hydrodynamic interactions with the gaseous halo.  For example,
several authors have suggested that the drag associated with motion
through the halo would absorb angular momentum from the fountain gas
\citep[e.g.,][]{fra08,mel08,mar11}.  This effect would increase the
tension between model predictions and our observations of high
specific angular momentum.  On the other hand, some models have
suggested that rapidly spinning halos could impart angular momentum to
ejected gas \citep[e.g.,][]{bro12}.  Furthermore, models of outflows
and inflows in some cosmological simulations suggest that most of the
\mgii\ gas in galaxy halos is associated with recycled accretion
\citep[e.g.,][]{for14}.  In that sense, we cannot rule out the
possibility that the gas we observe at $r\approx30$~kpc could have
previously been ejected in a wind.  That said, it is unclear whether
the specific angular momentum of ejected gas could be increased by a
factor of six to match our observations.  A more definitive answer
would require better understanding of the kinematics of hot halo gas
and its hydrodynamic interactions with ejected gas
\citep[e.g.,][]{bar06}.

\subsection{Comparison to models of cold-mode accretion}\label{sec:inflow}

While there is still debate about the fraction of gas accretion onto
galaxies that is shock heated to the virial temperature of the halo
\citep[e.g.,][]{whi91,ker05,nel13}, cosmological hydrodynamic
simulations generically predict that there should significant infall
of $10^{4}$--$10^{5}$~K gas onto galaxies, particularly for halo
masses $M_h\leq10^{12}$~\msun.  The filamentary structures that
provide gas to Milky Way--mass galaxies at $z\leq1$ are expected to be
less dense than filaments at $z\geq2$ and less likely to survive the
journey inwards from the virial radius to the outskirts of the
galactic disk \citep[e.g.,][]{ker09a}.  Nevertheless, some simulations
suggest that a fraction of the infalling cool gas at $z\leq1$ may
reach the outer disk and bring with it a significant amount of angular
momentum \citep[e.g.,][]{ros10,ste13,dan15}.  As discussed in
Section~\ref{sec:diswarp}, the existence of warped disks in the local
universe has been interpreted as evidence for cosmic infall, so it is
worth considering how the kinematics we observe compare with
predictions for infalling gas from zoom-in simulations.

One relevant comparison is with \citet{ste11a,ste11b}, who suggest
that the end of cold-mode accretion at $z\leq1$ for
$M_h\sim10^{12}$~\msun\ halos could be observed via \mgii\ absorption
systems that co-rotate with the galactic disk.  When tracking the
angular momentum of infalling gas in their simulations, they find that
recently accreted cool gas has a spin parameter $\approx4\times$
larger than that of the dark matter.  In their simulations, the most
recently accreted gas has the highest specific angular momentum and
forms an extended, warped disk with velocities that are offset from
system by $\pm100$--300~\kms.  The gas is expected to sink towards the
center of the halo within 1--2 dynamical times \citep[1--2~Gyr at
  $z\approx1$;][]{ste13}.  While the interactions between outflows and
inflows of hot and cool gas are not yet well understood, the
properties of the cold flow disks in these simulations are broadly
consistent with our observations.

\subsection{Comparison with high-velocity clouds, tidal streams, and extraplanar gas}\label{sec:disweak}

In addition to the dominant component of neutral gas in an extended
disk, deep \hi\ observations have revealed that galaxy halos in the
local universe often contain $10^{7}$--$10^9$~\msun\ of neutral gas
\citep[e.g.,][]{boo05,oos07,san08}.  This extraplanar halo gas has
been characterized in the Milky Way through the study of high-velocity
clouds \citep[HVCs; e.g.,][]{wak97,put12}, which trace
$M_{\scriptsize{\hi}}\approx3\times10^{8}$~\msun\ of neutral gas
associated with the Magellanic Stream, Leading Arm, and Magellanic
Bridge, plus an additional
$M_{\scriptsize{\hi}}\approx3\times10^{7}$~\msun\ distributed across
the entire sky \citep[e.g.,][]{put12}.  The typical mass of an HVC
complex is
$M_{\scriptsize{\hi}}=10^{5}$--$10^{6}$~\msun\ \citep{wak97,put12},
and cloud complexes as small as $M_{\scriptsize{\hi}}\approx10^{5}$
have been found in deep \hi\ observations of M31
\citep[e.g.,][]{thi04,wes07}.

The most prominent gaseous feature in the halo of the Milky Way is the
Magellanic Stream \citep[e.g.,][]{wan72,mat74}, which extends more
than 100$^{\circ}$ across the sky and contains more than
$10^8$~\msun\ of neutral gas \citep[e.g.,][]{put03,nid10}.  The
velocity offset between the Magellanic System and the Milky Way
\citep[$v_{\textnormal{\scriptsize{LMC}}}\approx380$~\kms;][]{kal06}
is comparable to what we observe for the optically thin component at
$v_{los}=-363$~\kms.  In addition to this systemic velocity offset,
the \feii\ column density and velocity width we find are comparable to
the observed properties of \feii\ absorbers in the Magellanic Stream
\citep[e.g.,][]{fox10}.  In that sense, the data are consistent with
this high-velocity component being associated with a Magellanic
Stream--like structure in the halo of the foreground galaxy.  We
visualize the potential location of such a stream in
Figure~\ref{fig:dis} with a blue square that is $d=50$~kpc away from
the center of the foreground galaxy, approximately equal to the
distance between the Milky Way and the Large Magellanic Cloud
\citep[e.g.,][]{fre10}.

The extraplanar halo gas observed within $\Delta z\sim10$~kpc of
nearby disk galaxies often shows evidence for co-rotation with the
thin disk, and there is evidence that gas at larger $z$ heights above
the disk tends to rotate more slowly
\citep[e.g.,][]{ran97,swa97,fra02}.  This extraplanar component that
has a connection to disk rotation is sometimes called ``disk-halo''
gas \citep[e.g.,][]{put12}.  In the context of this paper, it is
conceivable that the optically thin absorber at
$v_{los}=-256$~\kms\ could be tracing an extraplanar gas cloud with
kinematics dominated by rotation.  For example, locations near $|
\mathbf{L} |=40$~kpc ($z=16$~kpc, $r=27$~kpc) have
$v_{rot}/v_{los}\leq1.1$, implying that this velocity component could
have a rotational velocity as small as $v_{rot}\approx270$~\kms, which
would be consistent with the disk rotational velocity.  This location
is marked with a green triangle in Figure~\ref{fig:dis}.  Such a
configuration would imply essentially no change in $v_{rot}$ with $z$
height up to $\Delta z\approx8$~kpc above the warped disk, which would
be in contrast with the $\approx$15~\kms~kpc$^{-1}$ decrease in
rotation velocity with $z$ height for nearby galaxies like NGC~891
\citep[e.g.,][]{hea06}.  That said, the precise location of this cloud
along the line of sight cannot be uniquely determined from the
existing data, so it is not possible to draw definitive conclusions
about the vertical structure of co-rotating halo gas.  Nevertheless,
this intermediate-velocity absorber does exhibit the general
characteristics of a co-rotating extraplanar cloud near the disk-halo
interface.

\subsection{Constraints on cloud sizes}

The fact that we are observing gas in absorption towards an extended
background source provides an important constraint on the transverse
size of the absorbing clouds.  Previous studies have estimated clouds
sizes based on photoionization modeling \citep[e.g.,][]{rig02,sim06}
and based on differences between spectra of multiply lensed quasars
\citep[e.g.,][]{rau99,pet00}.  These studies often find cloud sizes
$l=N_H/n_H\approx100$~pc \citep[e.g.,][]{sch07,pro09,cri15} although
sizes estimates as small as $l\sim10$~pc \citep[e.g.,][]{rau99,rig02}
and as large as $l=1$--10~kpc \citep[e.g.,][]{sto13,leh13,wer14} are
not uncommon.

All three velocity components we observe are well fit by models with
unity covering fraction, implying that the absorbing clouds cover the
entire background source.  This is clear for the strong, saturated
\mgii\ and \feii\ lines at $v=-195$~\kms\ for which there is no
residual flux at line center.  For the weaker components at
$v=-256$~\kms\ and $v=-363$~\kms, the depth of the \mgii~$\lambda2796$
lines ($\tau\approx2$) approach zero, and the relative depth of the
\mgii~$\lambda2803$ lines ($\tau\approx1$) confirms that the
\mgii~$\lambda2796$ lines are not highly saturated.  In other words,
the small residual flux at line center for \mgii~$\lambda2796$ in
these weaker components is explained by a moderate column density
($N_{\scriptsize{\mgii}}\approx10^{13}$~cm$^{-2}$) with unity covering
factor rather than a higher column density with partial covering of
the background source.

Regarding the size of the background source, we use the radius that
contains 90\% of the total flux ($r_{90}$).  Based on the best-fit de
Vaucouleurs profile \citep{dia12,gea14}, this corresponds to
$r_{90}=520$~pc \citep[$r_{90}/r_{50}=5.5$;][]{gra05}.  Accounting for
the difference in angular diameter distance between $z_{bg}=0.712$ and
$z_{fg}=0.413$, the relevant size scale at the redshift of the
foreground galaxy is $l=390$~pc.  Using this as an estimate of the
size of the absorbing clouds, we can calculate the volume density of
the gas ($n_H=N_{H}/l$) and the cloud mass ($M_c=4/3\pi(l/2)^3 n_H m_p
\mu$) from the estimated column density.
\begin{equation}\label{eqn:density}
n_H = 0.01~\textnormal{cm}^{-3}\left(\frac{N_H}{1.2\times10^{19}~\textnormal{cm}^{-2}}\right)\left(\frac{390~\textnormal{pc}}{l}\right)
\end{equation}
\begin{equation}\label{eqn:mass}
M_c = 8~\msun\left(\frac{N_H}{1.2\times10^{19}~\textnormal{cm}^{-2}}\right)\left(\frac{l}{390~\textnormal{pc}}\right)^2
\end{equation}
The fiducial value $N_H=1.2\times10^{19}$~cm$^{-2}$ is for the
strongest component, assuming no dust depletion and solar metallicity.
In comparison, the $N_H$, $n_H$, and $M_c$ estimates for the weaker
components are smaller by a factor of $30$--40.  As discussed in
Section~\ref{sec:nh}, these $N_H$ values would be larger if the
gas-phase abundance of \feii\ were reduced by dust depletion or
sub-solar metallicity.

The most robust constraint on cloud properties associated with our
analysis is the $l=390$~pc lower limit on the transverse cloud size.
We have assumed in equations~\ref{eqn:density} and \ref{eqn:mass} that
this value for the transverse size is comparable to the absorption
path length, which may or may not be accurate.  Our size estimate is
consistent with the range of cloud sizes found in previous studies,
although there is some tension with smaller size estimates found
previously for weak \mgii\ absorbers \citep[$l\sim10$~pc;
][]{rau99,rig02}.  Our results suggest that low-ionization clouds with
densities as small as $n_{H}=3\times10^{-4}$~cm$^{-3}$ and masses as
small as $M_c=0.2$~\msun\ can survive in galaxy halos at distances
$r=30$--$50$~kpc.

\subsection{Implications for the geometry and kinematics of circumgalactic gas}

A number of previous studies have combined high-resolution
spectroscopy of background sources with low-resolution spectroscopy of
foreground galaxies to test kinematic models of extended disks,
inflows, and outflows \citep[e.g.,][]{ste02,kac10,bou13,bur13,che14}.
There are clear examples where individual absorption components can be
associated with an extended disk \citep[e.g.,][]{ste02,kac10,kac11b},
but pure disk models can rarely explain all of the absorption
components.  This is consistent with earlier work on the kinematics of
\mgii\ absorbers, which found that the data could be described by a
combination of stronger ``disk'' and weaker ``halo'' components
\citep[e.g.,][]{bri85,lan92,cha98}.

In this paper, we build on this previous work by expanding the
modeling to include warped disks, which are known to be ubiquitous
among disk galaxies in the local universe.  The existence of warps has
implications for estimates of the galactocentric distance and
three-dimensional kinematics associated with extended disks, and for
the range of azimuthal angles over which disk-related absorption could
be detected.  In addition, we show how individual absorption
components that are not associated with an extended disk can be
associated with other structures that are known to exist in galaxy
halos, including extraplanar gas and tidal streams.  Furthermore, we
demonstrate how this science can be done using background galaxies
rather than background quasars, which provides complementary
information about the sizes, densities, and masses of the absorbing
components.

\section{Summary}

We have used high-resolution spectroscopy of an extended background
source to study the circumgalactic gas associated with a Milky
Way--like galaxy at $z=0.413$.  We find evidence for an
$r\approx30$~kpc warped disk that follows the rotation curve of the
foreground galaxy, consistent with observations of extended \hi\ disks
in the local universe (see Figure~\ref{fig:fg_hires},
Sections~\ref{sec:warp}, Section~\ref{sec:diswarp}).  In the context
of theoretical models, this gas could be associated with recycled
accretion that has acquired a significant amount of angular momentum
since being ejected $>1$~Gyr ago (Section~\ref{sec:fountain}), and it
is also consistent with predictions for infalling cool gas that
co-rotates with the galactic disk (Section~\ref{sec:inflow}).  In
principle, a large-scale galactic wind with half-opening angle
$\theta_w\approx50^{\circ}$ and $v_w\approx500$~\kms\ could reproduce
similar line-of-sight kinematics, but the low $\sigmasfr$ for the
galaxy and the narrow line-of-sight velocity range for the absorbers
suggest that this is unlikely (see Sections~\ref{sec:outflow},
\ref{sec:disoutflow}).  We also detect two weak \mgii\ absorbers whose
kinematics are consistent with being associated with a tidal stream
and co-rotating extraplanar gas (see Section~\ref{sec:disweak},
Figure~\ref{fig:dis}).  The extended nature of the background source
allows us to place a lower limit on the transverse size of the
absorbing clouds ($l>0.4$~kpc), which is broadly consistent previous
estimates in the literature.

\acknowledgments

We acknowledge useful discussions with and assistance from James Aird,
James Bullock, John Chisholm, Jay Gallagher, Du{\v s}an Kere{\v s},
Britt Lundgren, Xavier Prochaska, Marc Rafelski, Kate Rubin, Jonathan
Whitmore, and Art Wolfe.  AMD acknowledges support from The Grainger
Foundation and from the Southern California Center for Galaxy
Evolution, a multi-campus research program funded by the University of
California Office of Research.  ALC acknowledges support from NSF
CAREER award AST-1055081.  RCH acknowledges support from an Alfred
P. Sloan Research Fellowship and a Dartmouth Class of 1962 Faculty
Fellowship.  Support for HST-GO-12272 was provided by NASA through a
grant from STScI.  Support for Spitzer-GO-60145 was provided by
contract 1419615 from JPL/Caltech.  Some of the data presented herein
were obtained at the W.M. Keck Observatory, which is operated as a
scientific partnership among the California Institute of Technology,
the University of California and the National Aeronautics and Space
Administration.  The Observatory was made possible by the generous
financial support of the W.~M. Keck Foundation.  The authors wish to
recognize and acknowledge the very significant cultural role and
reverence that the summit of Mauna Kea has always had within the
indigenous Hawaiian community.  We are most fortunate to have the
opportunity to conduct observations from this mountain.

\begin{center}
\begin{deluxetable}{cclcccrrrrr}
\tabletypesize{\scriptsize}

\tablecaption{Parameters of model fits to circumgalactic absorption lines\label{tab:foreground}}
\tablewidth{0pt}

\tablehead{ \colhead{$z$} & \colhead{$v$} & \colhead{ion} &
  \colhead{$\log{N}$} & \colhead{$b$} & \colhead{line} &
  \colhead{$\tau$} & \colhead{EW} & \colhead{line} & \colhead{$\tau$}
  & \colhead{EW} }

\startdata
$0.412494\pm0.000002$ & $-194.5\pm0.4$ & \mgii & $14.52\pm0.12$ & $23.9\pm1.2$ & 2796.354 & 35.72 & 0.90 & 2803.531 & 17.87 & 0.82 \\ 
                      &                & \feii & $14.62\pm0.05$ & $24.2\pm0.9$ & 2344.213 & 6.86  & 0.58 & 2374.460 & 1.91  & 0.37 \\
                      &                &       &                &              & 2382.764 & 19.58 & 0.71 & 2586.650 & 4.59  & 0.58 \\
                      &                &       &                &              & 2600.172 & 15.96 & 0.75 &      &       &      \\
$0.412206\pm0.000004$ & $-255.5\pm0.8$ & \mgii & $13.03\pm0.04$ & $12.7\pm1.1$ & 2796.354 & 2.17  & 0.24 & 2803.531 & 1.09  & 0.16 \\ 
                      &                & \feii & $13.13\pm0.05$ & $14.1\pm1.9$ & 2344.213 & 0.39  & 0.07 & 2374.460 & 0.11  & 0.02 \\ 
                      &                &       &                &              & 2382.764 & 1.10  & 0.15 & 2586.650 & 0.26  & 0.05 \\
                      &                &       &                &              & 2600.172 & 0.90  & 0.15 &      &       &      \\ 
$0.411700\pm0.000002$ & $-362.9\pm0.5$ & \mgii & $12.89\pm0.03$ & $11.6\pm0.9$ & 2796.354 & 1.70  & 0.20 & 2803.531 & 0.85  & 0.12 \\ 
                      &                & \feii & $13.03\pm0.04$ & $8.9\pm1.2$  & 2344.213 & 0.48  & 0.05 & 2374.460 & 0.13  & 0.02 \\ 
                      &                &       &                &              & 2382.764 & 1.35  & 0.11 & 2586.650 & 0.32  & 0.04 \\
                      &                &       &                &              & 2600.172 & 1.10  & 0.11 &      &       &      \\ 
\enddata

\tablecomments{Col. (1): Redshift.  Col. (2): Velocity [\kms].
  Col. (3): Ion.  Col. (4): Log column density [cm$^{-2}$].  Col. (5):
  Doppler parameter [\kms].  Col. (6--11): Transitions considered,
  along with optical depth at line center and rest-frame equivalent
  width [\AA] for each transition.}
\end{deluxetable}
\end{center}

\end{document}